\documentclass[aps, pop, superscriptaddress, twocolumn, showpacs]{revtex4}
\usepackage{graphicx}    
\usepackage{bm}          

\usepackage{color}
\usepackage{CJK}
\begin{document}
\title{Simulation of Stimulated Brillouin Scattering and Stimulated Raman Scattering In Shock Ignition}

\author{L. Hao\footnote{Permanent Address: Institute of Applied Physics and Computational Mathematics, Beijing, 100094, China}}
\affiliation{Department of Mechanical Engineering and Laboratory for
Laser Energetics, University of Rochester, Rochester, New York
14627, USA}

\author{J. Li}
\affiliation{Department of Mechanical Engineering and Laboratory for Laser Energetics, University of Rochester,
Rochester, New York 14627, USA}

\author{W. D. Liu}
\affiliation{Department of Mechanical Engineering and Laboratory for Laser Energetics, University of Rochester,
Rochester, New York 14627, USA}

\author{R. Yan}
\affiliation{Department of Mechanical Engineering and Laboratory for Laser Energetics, University of Rochester,
Rochester, New York 14627, USA}

\author{C. Ren\footnote{Email: chuang.ren@rochester.edu}}

\affiliation{Department of Mechanical Engineering and Laboratory for Laser Energetics, University of Rochester,
Rochester, New York 14627, USA}
\affiliation{Department of Physics and Astronomy, University of Rochester, Rochester, New York 14627, USA}

\date{\today}

\begin{abstract}
We study stimulated Brillouin scattering (SBS) and stimulated Raman
scattering (SRS) in shock ignition by comparing fluid and PIC
simulations. Under typical parameters for the OMEGA experiments
[Theobald \emph{et al}., Phys. Plasmas \textbf{19}, 102706 (2012)],
a series of 1D fluid simulations with laser intensities ranging
between 2$\times 10^{15}$ and 2$\times 10^{16}$ W/cm$^2$ finds that
SBS is the dominant instability, which increases significantly with
the incident intensity. Strong pump depletion caused by SBS and SRS
limits the transmitted intensity at the 0.17$n_c$ to be less than
3.5$\times 10^{15}$ W/cm$^2$. The PIC simulations show similar
physics but with higher saturation levels for SBS and SRS convective
modes and stronger pump depletion due to higher seed levels for the
electromagnetic fields in PIC codes. Plasma flow profiles are found
to be important in proper modeling of SBS and limiting its
reflectivity in both the fluid and PIC simulations.
\end{abstract}

\maketitle

\section{Introduction}

Shock ignition (SI) \cite{Betti07} is a new high gain ignition
scheme in inertial confinement fusion (ICF). The key process in SI
is to generate a strong shock at the end of the compression stage by
escalating the intensity of the incident laser to
$10^{15}\sim10^{16}$W/cm$^2$ \cite{Betti07,Perkins09}. In this high
laser intensity regime, growth rates of many laser plasma
instabilities (LPI) exceed their thresholds, such as the two plasmon
decay (TPD), the stimulated Raman scattering (SRS), and the
stimulated Brillouin scattering (SBS). Previous Particle-in-Cell
(PIC) simulations showed large laser reflectivities at high
intensities due to SBS and SRS \cite{Klimo10,Riconda11}, saturation
of TPD due to plasma cavity formation \cite{Weber12}, intermittent
LPI activities due to interplay of modes at different density
regions \cite{Yan14}, and LPI's dependence on plasma temperatures
\cite{Weber15}. Both the integrated SI experiment on OMEGA
\cite{Theobald12} and the PIC simulation \cite{Yan14} showed that
the LPI generated hot electrons with a temperature of $\sim 30$ keV.
A recent short pulse experiment found high level reflectivity of SBS
under SI-relevant intensities \cite{Goyon13}. Due to the sensitivity
and nonlinearity of LPI's dependence on laser and plasma conditions,
it is very important to explore the wide parameter space and
understand the physics among the complicated LPI instabilities in
SI.

In the low density region of the corona, SBS and SRS can be
convective and their saturation levels depend on their seed levels
as well as their convective gains. It is well known that in PIC
codes high frequency modes of the electromagnetic fields have much
higher noise levels than an actual plasma of the same physical
conditions\cite{okuda72}. How the inflated seed levels affect LPI in
SI has not been studied so far. In addition the previous PIC
simulations \cite{Klimo10,Riconda11,Weber12,Yan14,Weber15} did not
include plasma flows, which can affect SBS reflectivity through the
detuning of the ion acoustic wave resonance \cite{Galeev73}. In this
paper, we study SBS and SRS for typical shock ignition conditions,
including the flow velocity gradient, via a series of fluid and PIC
simulations for the first time. Our fluid simulation results show
that SBS is the dominant cause of the strong pump depletion for
laser intensities of $I=2\times 10^{15} \sim 2\times 10^{16}$ W/cm
$^2$, and the flow velocity gradient has an important effect on
limiting the SBS reflectivity. The transmitted laser intensity near
the quarter critical region is limited to an asymptotic value of
$I\sim 3.3\times 10^{15}$ W/cm $^2$, which should be taken into
account in SI design. The PIC simulations show similar physical
trends as the fluid results but stronger pump depletion and higher
saturation levels of SBS and SRS due to the higher numerical seed
levels. The results here show the importance of incorporating
realistic seed levels for correctly modeling LPI in the SI regime.

\section{The simulation results}

Both the fluid and PIC simulations have been performed in one
dimension (1D) to study SBS and SRS in the low density region before
the quarter critical surface. They are complementary to the 2D
simulations in Ref. \cite{Yan14} where the lowest density was
$n=0.17 n_c$ ($n_c$ is the critical density) and TPD was also
studied. The PIC simulations are performed with OSIRIS
\cite{Fonseca02}. The fluid simulations are performed with the HLIP
code \cite{Hao14}, which is a 1D steady-state code solving the
coexistent problem of SBS and SRS along the ray path, similar to
DEPLETE \cite{Strozzi08}. The main equations in HLIP are listed as
follows.
\begin{eqnarray}\label{e1}
\frac{{\partial {I_0}}}{{\partial z}} =  - \frac{{2{\nu
_0}}}{{{v_{g0}}}}{I_0} - {\omega _0}{I_0}\sum\limits_{s=R,B} {\int
{\frac{{({K_s}{Z_s} + {T_s})}}{{{\omega _s}}}d} {\omega _s}},
\end{eqnarray}
\begin{eqnarray}\label{e2}
\frac{{\partial {Z_s}}}{{\partial z}} = \frac{{2{\nu
_s}}}{{{v_{gs}}}}{Z_s} - {K_s}{I_0}{Z_s} - {I_0}{T_s},(s=R, B)
\end{eqnarray}
where $I$, $\omega$, $\nu$ and $v_g$ denote the laser intensity, the
angular frequency, the collisional damping rate and the group
velocity, respectively. The subscript $0$ refers to the incident
light, and the subscript $s$ refers to the backscattered light from
either SBS or SRS. Here, $Z_s$ denotes the intensity per angular
frequency of the backscattered light with the integrated intensity
${I_s}(z) = \int {{Z_s}({\omega_s},z)d{\omega_s}}$. The seed term
for the backscattered light $T_s$ is a calculated according to the
Thomson scattering model \cite{Berger89}. The coupling coefficient
\begin{eqnarray}\label{e3}
{K_s} = \frac{{2\pi (k_s+k_0)^2{e^2}}}{{{k_s}{k_0}{\omega
_0}m_e^2{c^4}}}{\mathop{\rm Im}\nolimits} \left[ \frac{\chi
_e}{\varepsilon}( 1 + \sum\limits_j \chi _j ) \right],
\end{eqnarray}
is the local spatial growth rate of the backscattered light, where
$k$ denotes the wavenumber of light, and as usual $e$, $m_e$, $c$
are the electron charge, the electron mass, and the speed of light
in vacuum, respectively. The susceptibility for the electrons and
the ion species $j$ are $\chi_e$ and $\chi_j$, respectively, and
$\varepsilon=1+\chi_e+\sum_j\chi_j$ is the dielectric function,
which depends on the frequency and wavenumber of the Langmuir wave
or ion acoustic wave driven by the ponderomotive force. The kinetic
term ${\rm Im}[\chi_e(1+\sum_j\chi_j)/\varepsilon]$ is the
ponderomotive response of the plasma to the light field
\cite{Drake74}, which contains the effect of both Landau damping and
phase detuning \cite{Strozzi08}.

\begin{figure}[htb!]
\includegraphics[height=0.25\textwidth,width=0.3\textwidth,angle=0]{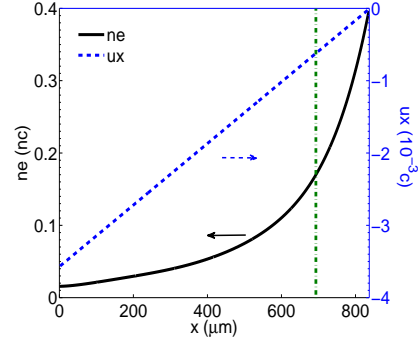}
\caption{(Color online) The normalized density profile (black solid
line) and plasma flow profile (blue dashed line) used in the simulations.} \label{fig1}
\end{figure}

Our simulation parameters are fitted from the
LILAC \cite{Delettrez} simulation results for the OMEGA integrated
SI experiments \cite{Theobald12}. The incident laser has a
wavelength of $\lambda_0=0.351\mu m$, and the length of the
simulation box is $L=836\mu$m. In Fig. \ref{fig1}, the black solid
line shows the density profile normalized by $n_c$ along the ray
path . It is the same profile used in the 1D simulations in Ref.
\cite{Yan14} and has a density scale length of $L_n=170 \mu$m at the
$1/4-n_c$ surface. The analytic expression of the normalized density
is $n_e(x)=a_6x^6+a_5x^5+a_4x^4+a_3x^3+a_2x^2+a_1x+a_0$, where $x$
is the longitudinal distance from left boundary of simulation box in
$\lambda_0$, $a_6=1.538\times 10^{-20},a_5=-8.225\times
10^{-17},a_4=1.797\times 10^{-13},a_3=-1.817\times 10^{-10},
a_2=9.523\times 10^{-8}, a_1=3.758\times 10^{-6}$, and $a_0=0.0156$.
The density range is from $0.0156 n_c$ to $0.4 n_c$. The blue dashed
line in Fig. \ref{fig1} shows the plasma flow profile normalized by
the vacuum speed of light $c$, in the form of
$u(x)=-0.003567+1.494\times 10^{-6}x$. The ion components in the
plasma are fully ionized C and H in 1:1 ratio. Two sets of the
plasma temperatures are chosen from the LILAC simulations. In the
low temperature (LT) case, $T_e=1.6$keV and $T_C=T_H=0.55$keV, which
corresponds to the temperatures at the launch of the ignition pulse.
In the high temperature (HT) case, $T_e=3.5$keV and
$T_C=T_H=1.6$keV, which represents the temperatures at the peak
intensity of the ignition pulse. A green dash-dot line is also drawn
in Fig. \ref{fig1} at the position of $n_e=0.17n_c$, where the
transmitted laser intensity is diagnosed in this paper.

\begin{figure}[htb!]
\includegraphics[height=0.35\textwidth,width=0.22\textwidth,angle=0]{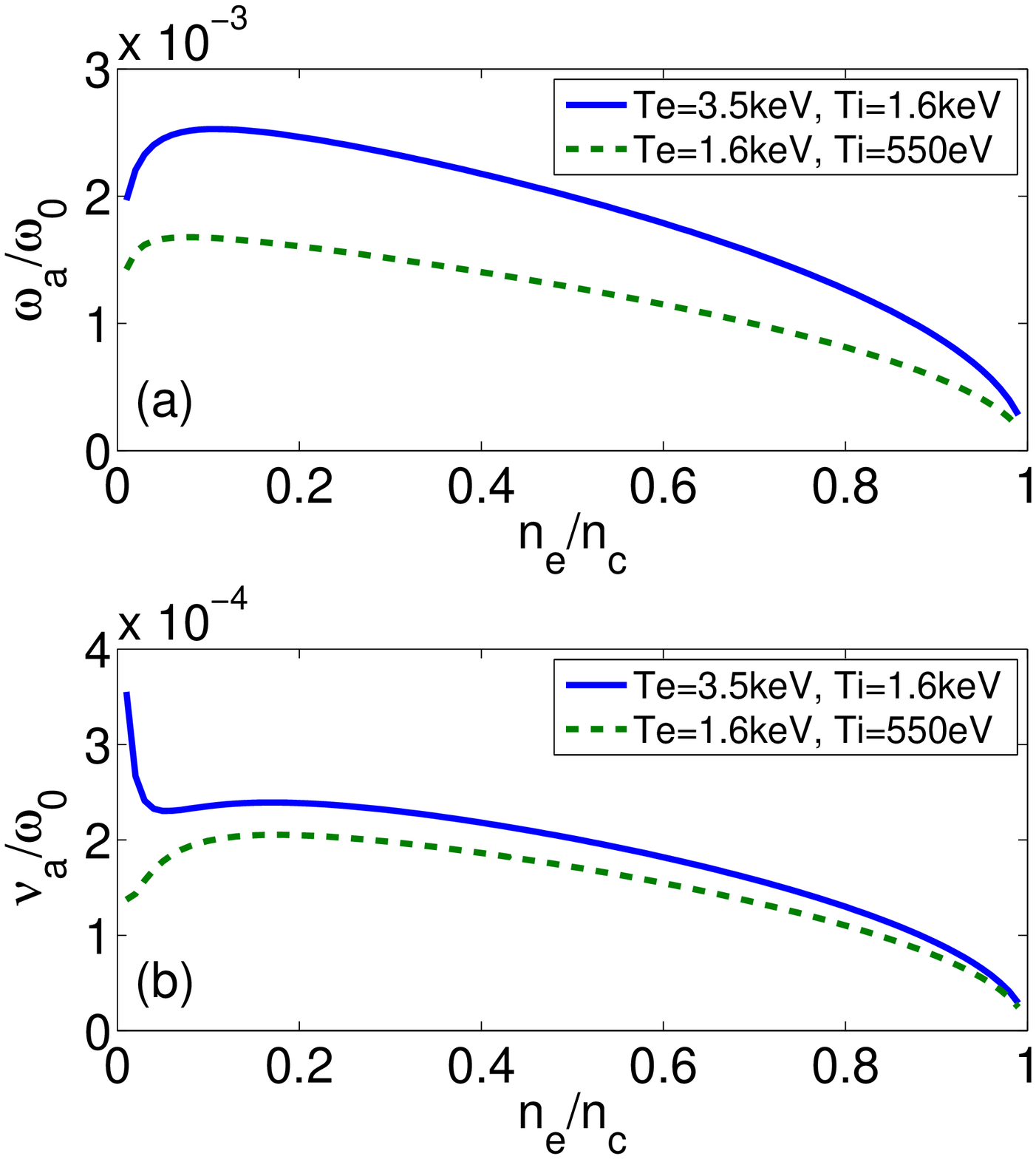}
\includegraphics[height=0.33\textwidth,width=0.25\textwidth,angle=0]{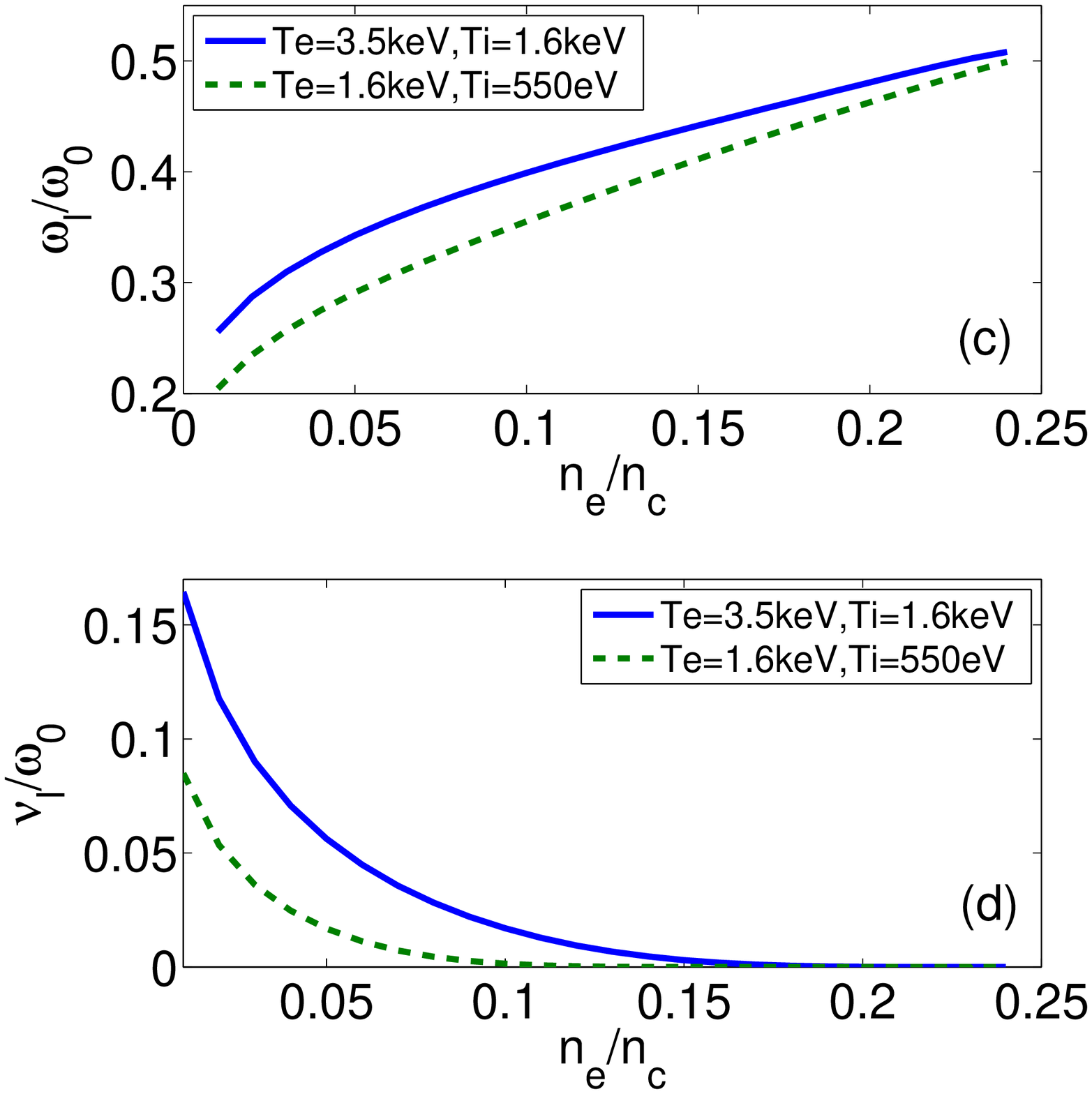}
\caption{(Color online) The frequency (a) and the Landau damping rate (b) of the least-damped ion-acoustic waves; and the frequency (c) and the Landau damping rate (d) of the least-damped Langmuir waves in the CH plasma under two different temperatures. } \label{fig2}
\end{figure}

For our CH plasma profile we have calculated the frequencies and
Landau damping rates of the least-damped ion-acoustic waves and
Langmuir waves for the two different temperature cases and the
results are plotted in Fig. \ref{fig2}. These results are obtained
by numerically solving the dispersion relation
$\varepsilon(\omega,k)=0$ combined with the matching condition of
the three-wave coupling
$(\omega_0-\omega)^2-\omega_{pe}^2=(k-k_0)^2c^2$. The results show
the ion-acoustic waves correspond to the weakly damped slow
ion-acoustic mode in Ref. \cite{vu94}. That for CH plasma the slow
ion-acoustic mode is dominant is also consistent with the conclusion
given by Williams \emph{et al.} \cite{Williams95}. The results also
show that the high temperature case has higher Landau damping rates
for both the ion acoustic waves and the Langmuir waves.

\subsection{The fluid simulation results}



\begin{figure}[htb!]
\includegraphics[height=0.45\textwidth,width=0.45\textwidth,angle=0]{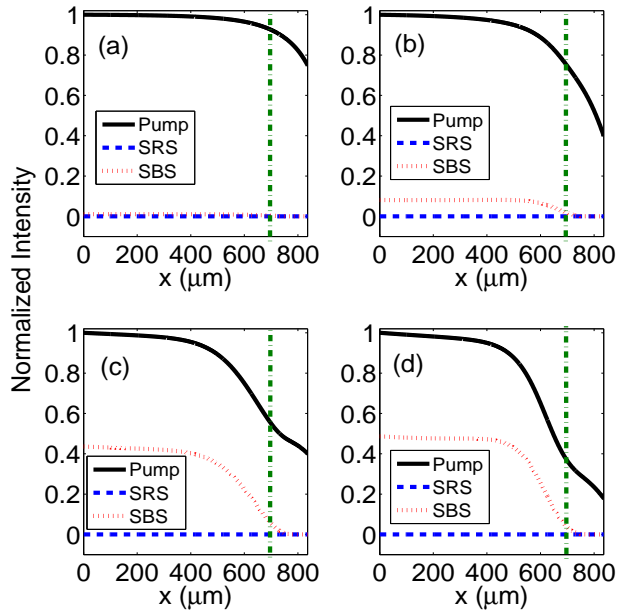}
\caption{(Color online) Spatial profiles of the normalized
intensities of the pump and backscattered light for the HT case
(a,c) and the LT case (b,d), both with the plasma flow. The incident
laser intensity is $I=2\times 10^{15}$W/cm$^2$ for (a,b) and
$I=5\times 10^{15}$W/cm$^2$ for (c,d).} \label{fig3}
\end{figure}

The fluid simulations for both the HT and LT cases have been
performed with the same plasma flow profile and different laser
intensities from $I=2\times 10^{15}$W/cm$^2$ up to $I=2 \times
10^{16}$ W/cm $^2$. The grid size is $\lambda_0$. Based on the
frequency range of the weakest damped modes of ion-acoustic wave and
Langmuir wave shown in Figs. \ref{fig2}(a) and \ref{fig2}(c), we set
the wavelength of backscattered light of SRS from $400$nm to
$715$nm, and the wavelength of backscattered light of SBS from
$350.5$nm to $353$nm in HLIP to include the weakest damped modes on
the ray path in our fluid simulations.

For each run, we can obtain the intensities of the pump laser and
backscattered light along the ray path in the steady-state. Parts of
the fluid simulation results are shown in Fig. \ref{fig3}, where the
black solid line, the red dotted line, and the blue dashed line
represent the intensity profiles of the pump laser, the SBS
backscattered light, and the SRS backscattered light, respectively.
All of the intensities are normalized by the incident laser
intensity, which is $I=2\times 10^{15}$W/cm$^2$ in Figs.
\ref{fig3}(a) and \ref{fig3}(b), and $I=5\times 10^{15}$W/cm$^2$ in
Figs. \ref{fig3}(c) and \ref{fig3}(d). They show that SBS is the
dominant instability, and the SBS reflectivity increases
significantly as the incident laser intensity becomes higher. The
SBS reflectivity is larger in the LT case than in the HT case,
because Landau damping of the ion-acoustic wave is weaker in the LT
case due to its larger $T_e/T_i\approx 2.9$ than the HT case
where $T_e/T_i\approx 2.2$, as shown in Fig. \ref{fig2}. 

\begin{figure}[htb!]
\includegraphics[height=0.23\textwidth,width=0.45\textwidth,angle=0]{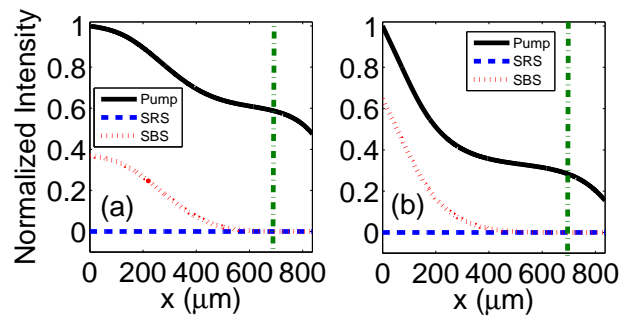}
\caption{(Color online) Spatial profiles of the normalized
intensities of the pump and backscattered light for (a) the HT case
and (b) the LT case, without the consideration of the plasma flow
$I=2\times 10^{15}$W/cm$^2$.} \label{fig4}
\end{figure}

In order to study the influence of the plasma flow on SBS, a set of
simulations at $I=2\times 10^{15}$W/cm$^2$ have also been performed
without the plasma flow, as shown in Figs. \ref{fig4}(a) and
\ref{fig4}(b) for the HT and LT cases, respectively. Comparing them
to Figs. \ref{fig3}(a) and \ref{fig3}(b), we can see that SBS is
reduced by the plasma flow, resulting in lower SBS reflectivities.
This is because the flow can Doppler shift the frequency of the
local ion-acoustic wave and the gradient of the flow velocity can
introduce phase mismatch between the SBS backscattered light wave
coming from the higher density region and the local ion acoustic
wave. This limits the further amplification of these convective
modes at the lower density region \cite{Galeev73}. 

The SRS reflectivity maintains at a low level ($<1\%$) in Figs.
\ref{fig3} and \ref{fig4}. Indeed, no considerable SRS reflectivity
is seen in the simulations until the laser intensity is higher than
$1.5\times 10^{16}$W/cm$^2$. That is because the density scale
length here, $L_n=170\mu$m, is small enough to detune the match
condition between the local electronic plasma wave and the SRS
backscattered wave from the higher density region. So the density
gradient limits the SRS reflectivity effectively in the lower laser
intensity cases \cite{Galeev73}.

\begin{figure}[htb!]
\includegraphics[height=0.25\textwidth,width=0.3\textwidth,angle=0]{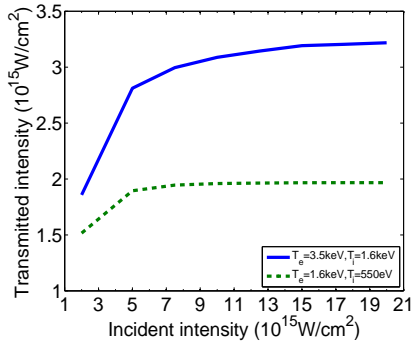}
\caption{(Color online) The transmitted laser intensity at different
incident laser intensity \textbf{with the consideration of plasma
flow}.} \label{fig5}
\end{figure}

\begin{figure}[htb!]
\includegraphics[height=0.25\textwidth,width=0.3\textwidth,angle=0]{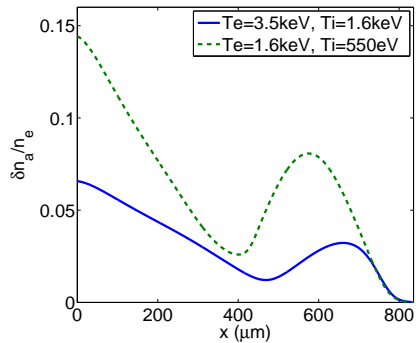}
\caption{(Color online) Amplitudes of electron density perturbations
due to SBS along the ray path.} \label{fig6}
\end{figure}

The transmitted laser intensity at $n_e=0.17n_c$ for different
incident laser intensity is shown in Fig. \ref{fig5}. It increases
with the incident laser intensity but tends to an asymptotic maximum
value, which is about 3.3$\times 10^{15}$ W/cm$^2$ for the HT case
and 2$\times 10^{15}$ W/cm$^2$ for the LT case, due to larger
reflectivity at high intensities. The transmitted intensity is
always lower in the LT case than in the HT case, because Landau
damping of the ion-acoustic wave and electronic plasma wave is lower
in the LT case, resulting in stronger SBS and SRS activities.
Although SBS is always the dominant instability in our fluid
simulations, the SRS reflectivity is also considerable ($<10\%$) for
the higher incident laser intensities. Fig. \ref{fig6} shows the
amplitudes of the electron density perturbations due to SBS at
$I=5\times 10^{15}$W/cm$^2$ for both the HT and LT cases, showing
$\delta n_a/n_e<0.15$, indicating nonlinear effects are not
important here. For $I=2\times 10^{16}$W/cm$^2$ case, $\delta
n_a<0.015n_c$ remains small even though $\delta n_a/n_e$ can be as
large as 0.9 at left boundary due to very small $n_e$. Nonlinear
effects at the high intensity may be important.

\subsection{The PIC simulation results}

The OSIRIS \cite{Fonseca02} simulations use the same plasma
parameters as shown in Fig. \ref{fig1}, and laser intensities of
$I=2, 5, 10\times 10^{15}$W/cm$^2$ for both the HT and LT cases. All
the simulations use the 2nd-order spline current deposition scheme
with current smoothing. The grid size is $\Delta x=0.1c/\omega_0$,
and the time step is $\Delta t=0.0707/\omega_0$. The electron-ion
collision is included in all PIC simulations by turning on the
binary collision module in OSIRIS \cite{Yan12}. Boundary conditions
are chosen to be open for the electromagnetic fields, and thermal
bath for the particles \cite{Yan14}. The flow velocity profile is
implemented by adding a flow velocity $u(x)$ to the thermal
velocities of particles initially with $u(L)=0$ at the right
boundary. At the left boundary $u(0)=-0.003567c$. The flow profile
has the same velocity gradient as the LILAC simulation. The
particles drift toward the left boundary and are re-injected into
the simulation box with the initial temperature. The difference in
the particle energy is recorded to diagnose the net energy flux of
the particles leaving the simulation box.

\begin{figure}[htb!]
\includegraphics[height=0.25\textwidth,width=0.4\textwidth,angle=0]{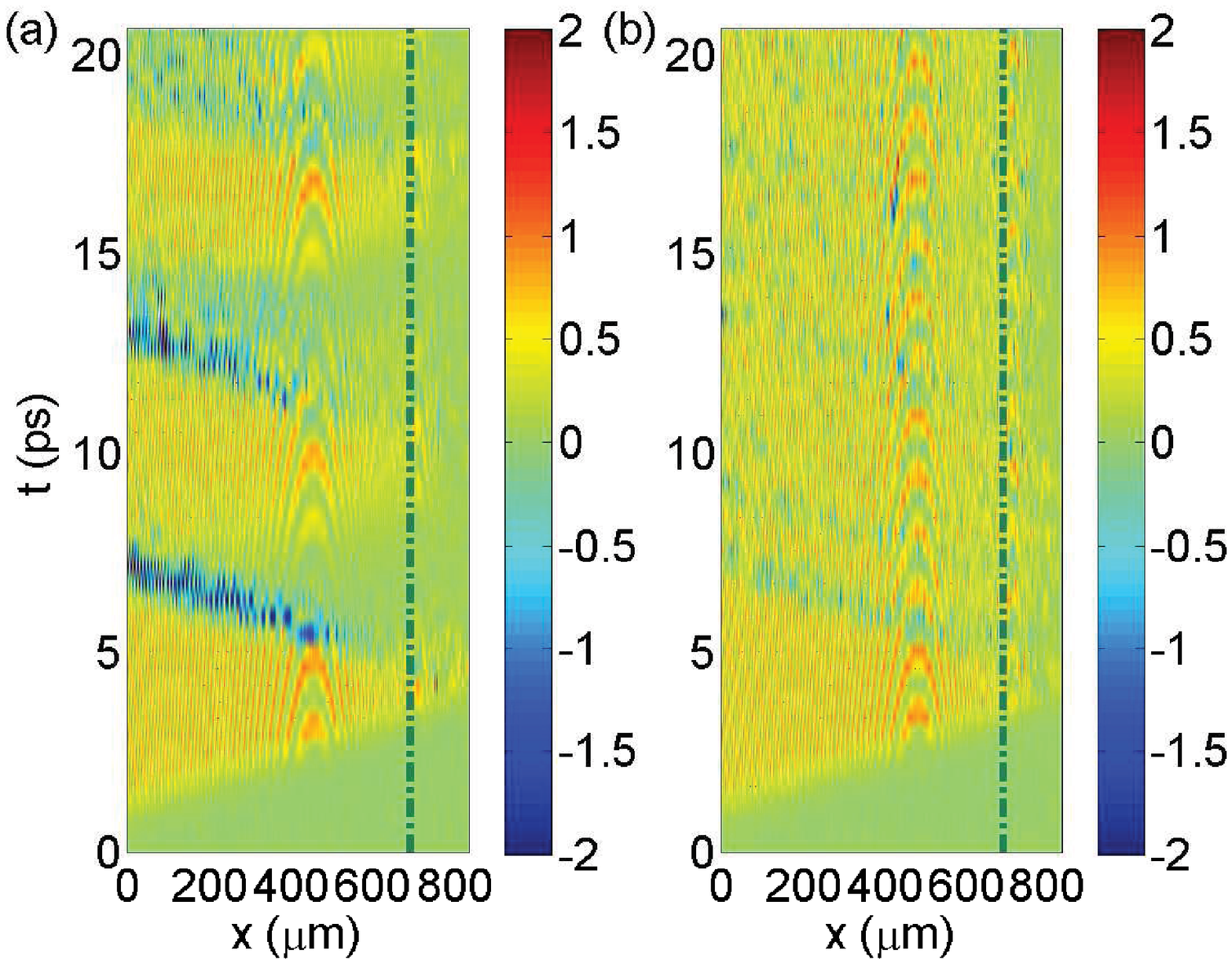}
\\\
\\\
\includegraphics[height=0.25\textwidth,width=0.4\textwidth,angle=0]{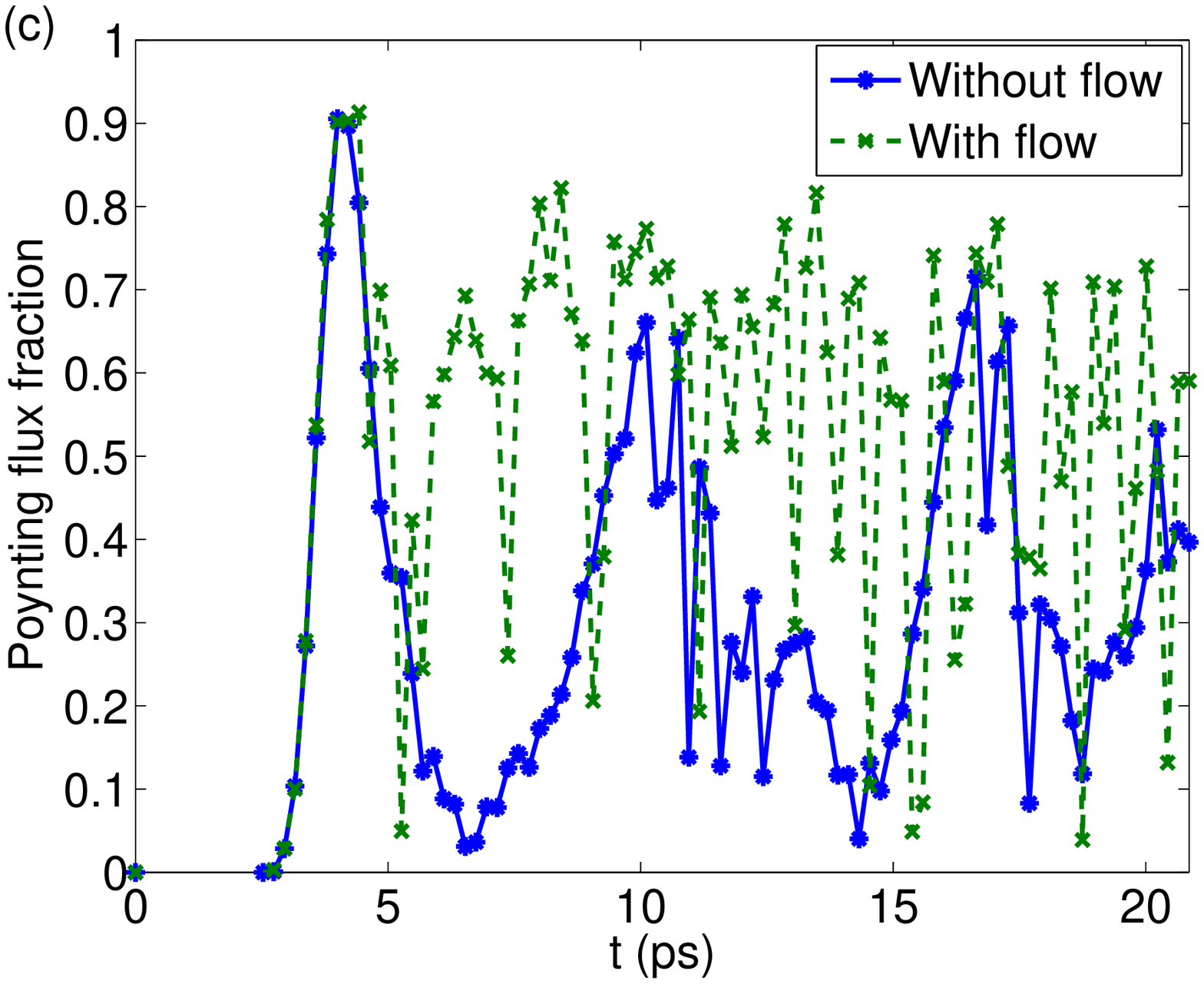}
\caption{(Color online) Evolution of the normalized Poynting vectors
for the HT case (a) without flow and (b) with flow, and (c) the time
resolved Poynting flux fraction at $n_e=0.17n_c$. The incident laser
intensity is $2\times 10^{15}$W/cm$^2$.} \label{fig7}
\end{figure}

Fig. \ref{fig7} shows the comparison of the PIC simulation results
with and without the plasma flow velocity in the HT case with
$I=2\times 10^{15}$W/cm$^2$. In Figs. \ref{fig7}(a) and
\ref{fig7}(b), the longitudinal component of the Poynting vector,
normalized by the incident Poynting vector at the left boundary, is
shown. Stronger bursts of backscattered light are seen in the case
without the plasma flow. This also indicates that the reflectivity
is largely due to SBS since SRS is not sensitive to the plasma flow
velocity. Correspondingly, the transmitted Poynting flux fraction at
$n_e=0.17n_c$ shows that the pump depletion is also stronger when
the plasma flow velocity is not considered, which is consistent with
our fluid simulations.

\begin{figure}[htb!]
\includegraphics[height=0.23\textwidth,width=0.23\textwidth,angle=0]{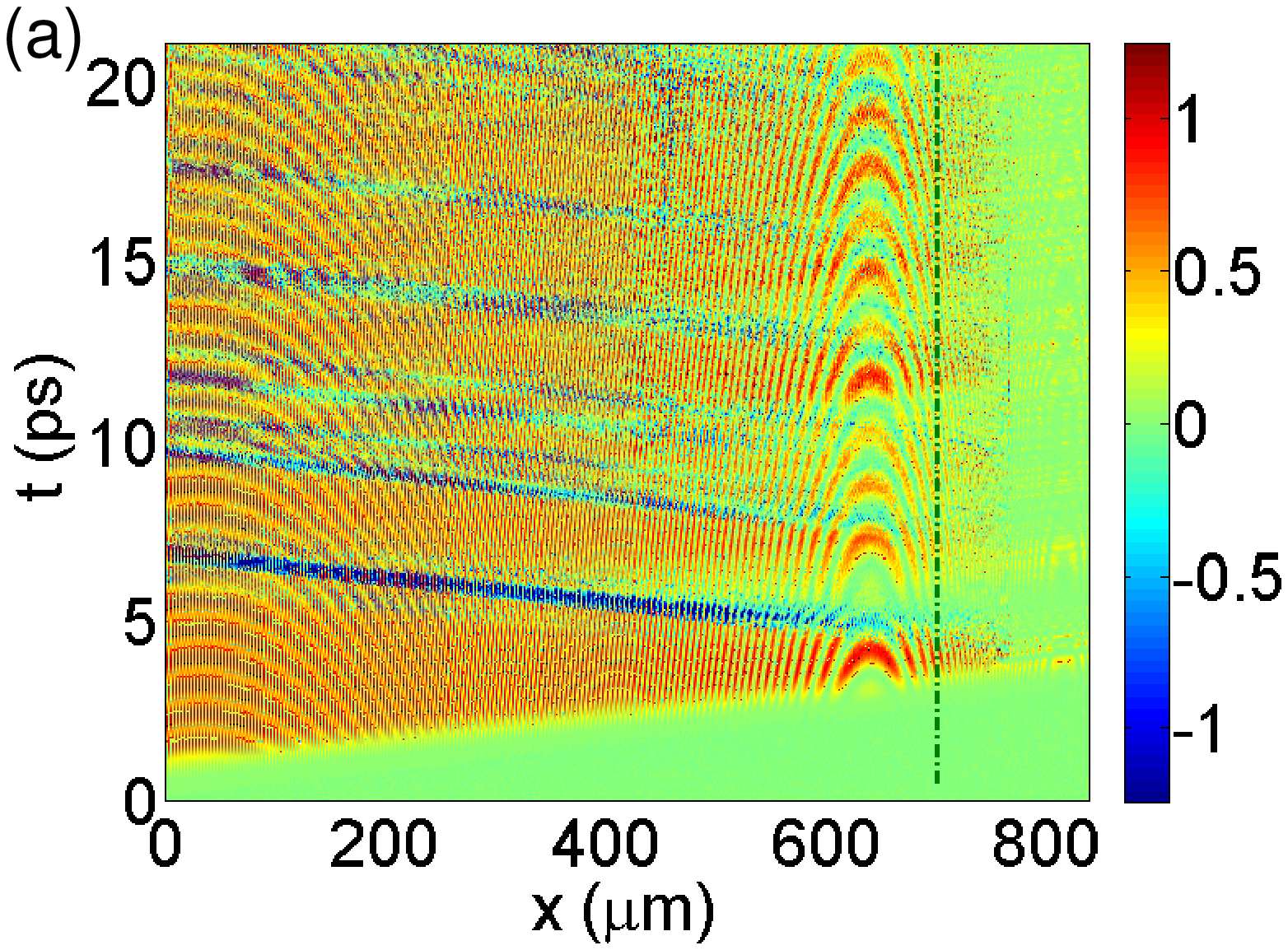}
\includegraphics[height=0.23\textwidth,width=0.23\textwidth,angle=0]{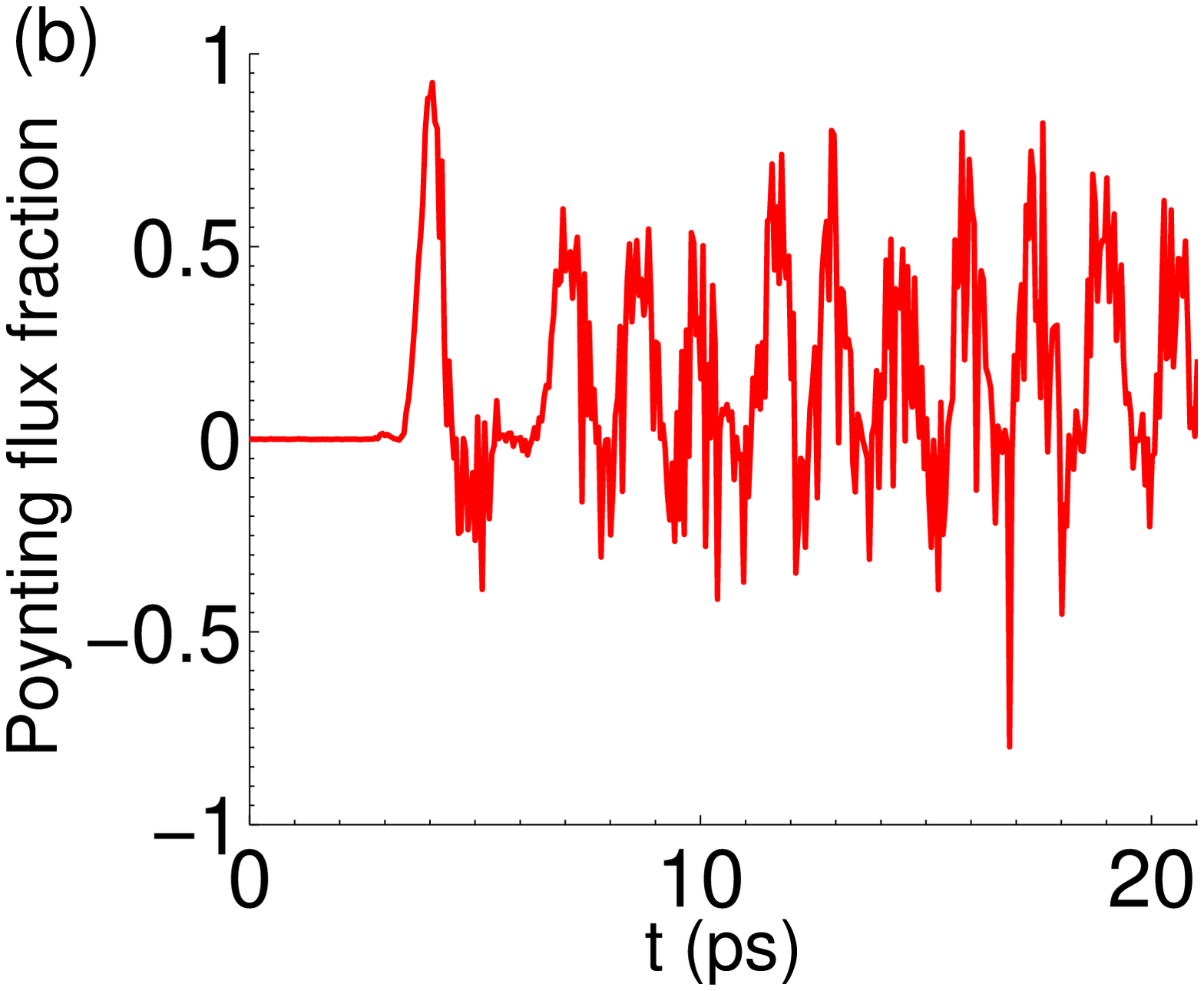}
\\\
\\\
\\\
\includegraphics[height=0.23\textwidth,width=0.23\textwidth,angle=0]{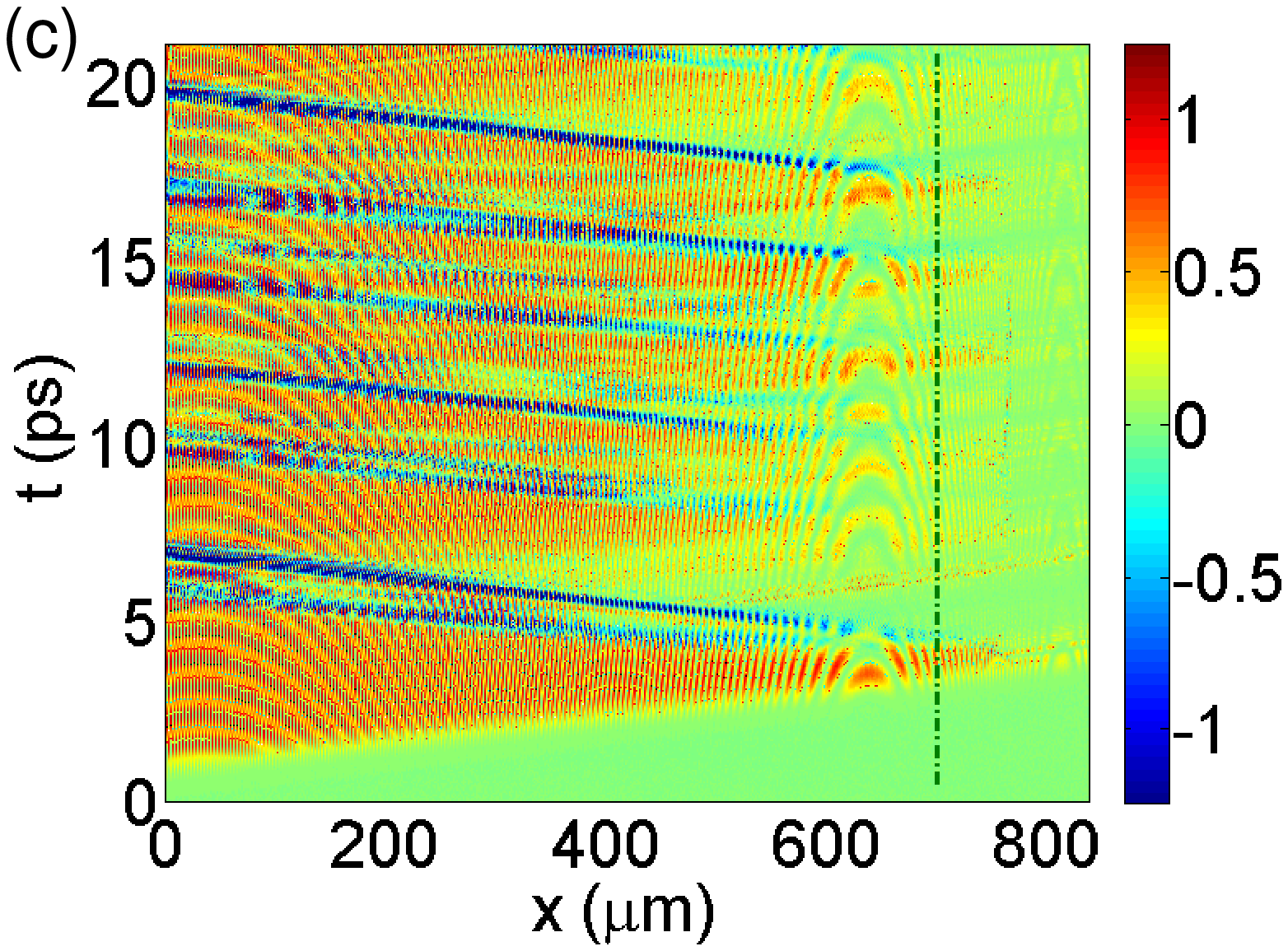}
\includegraphics[height=0.23\textwidth,width=0.23\textwidth,angle=0]{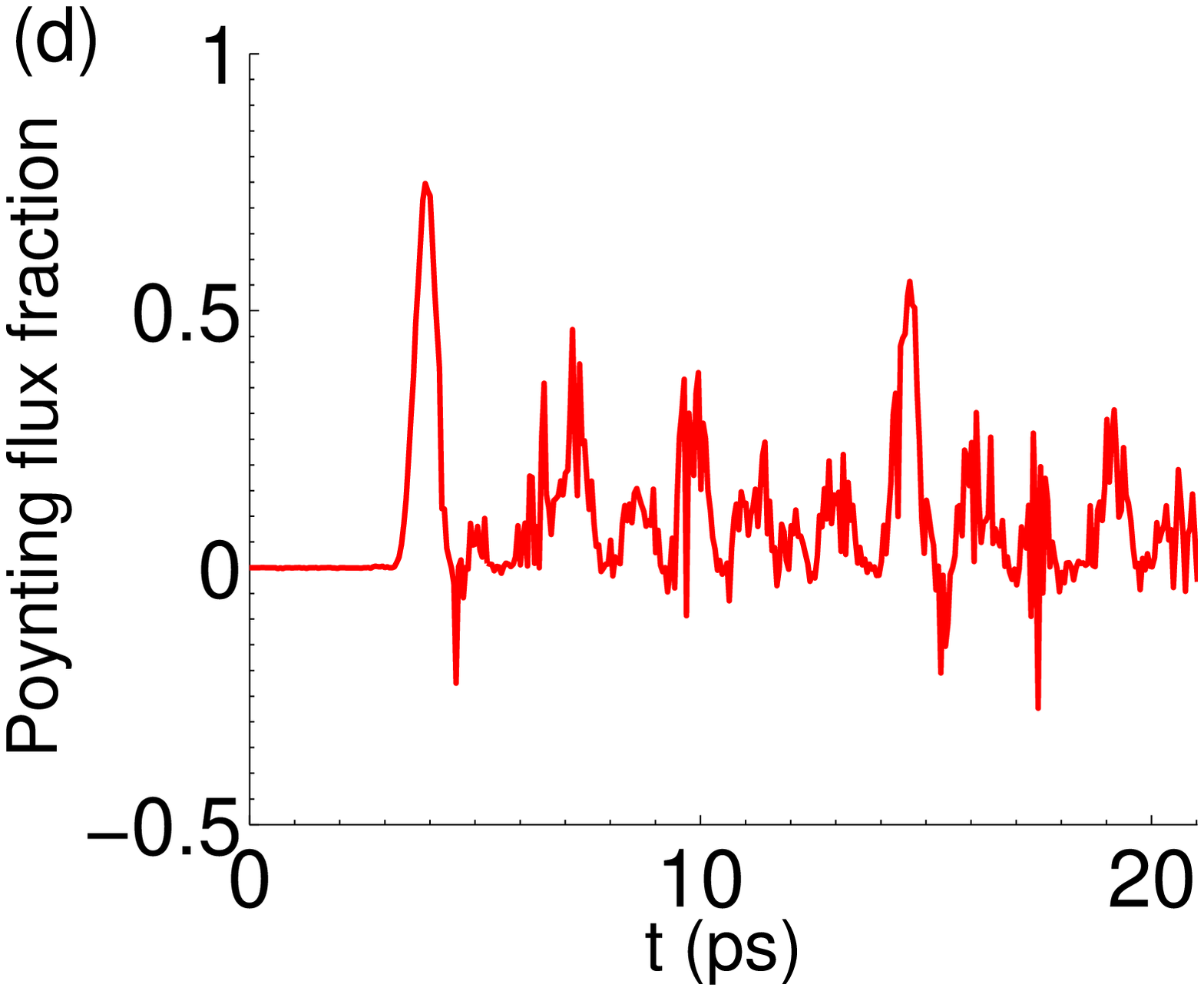}
\caption{(Color online) Evolution of normalized Poynting vectors for
(a) the HT case and (c) the LT case, and the time resolved Poynting
flux fraction at $n_e=0.17n_c$ for (b) the HT case and (d) the LT
case. The incident laser intensity is $5\times 10^{15}$W/cm$^2$ and
the plasma flow is included.} \label{fig8}
\end{figure}

Figs. \ref{fig8}(a) and \ref{fig8}(c) show the evolution of the
Poynting vectors for the HT and LT cases respectively, when
$I=5\times 10^{15}$W/cm$^2$ and with the plasma flow. To separate
SBS and SRS reflectivities in the PIC simulations, we spatially
Fourier-transform the $B_z$ field near the left boundary, and filter
the data within the wavenumber region of SRS backscattered light in
k-space to obtain the SRS reflectivity at every dumping step. The
temporal average value for the total reflectivity can be obtained
from the Poynting vector at left boundary. The SBS reflectivity is
the difference of the two. We find a temporal average SBS
reflectivity of approximately $31\%$ in the HT case and $51\%$ in
the LT case, which indicates that SBS is also strong in the PIC
simulations, especially for the LT case. The time resolved
transmitted Poynting flux fraction at $n_e=0.17n_c$ is shown in
Figs. \ref{fig8}(b) and \ref{fig8}(d). Its temporal average value is
$20\%$ in the LT case, lower than the $32\%$ in the HT case. This
indicates a significant SRS reflectivity, which is different from
the fluid results.

\begin{table}
 \caption{Simulated transmitted intensity fractions}
  \begin{center}
    \begin{tabular}{c c c c}
      \hline
      \hline
      Laser intensity & Temperature & \multicolumn{2}{c}{Transmitted intensity fraction} \\ \cline{3-4}
      (W/cm$^2$)& conditions & HLIP & OSIRIS \\ \hline
      $2\times10^{15}$ & HT case & $93\%$ & $60\%$ \\
        & LT case & $76\%$ & $25\%$ \\
      \hline
      $5\times10^{15}$ & HT case & $56\%$ & $32\%$ \\
        & LT case & $38\%$ & $20\%$ \\
      \hline
      $1\times10^{16}$ & HT case & $31\%$ & $12\%$ \\
         & LT case & $20\%$ & $8\%$ \\
      \hline
      \hline
    \end{tabular}
  \end{center}\label{table1}
\end{table}

\subsection{The seed level analysis}

\begin{figure}[htb!]
\includegraphics[height=0.25\textwidth,width=0.4\textwidth,angle=0]{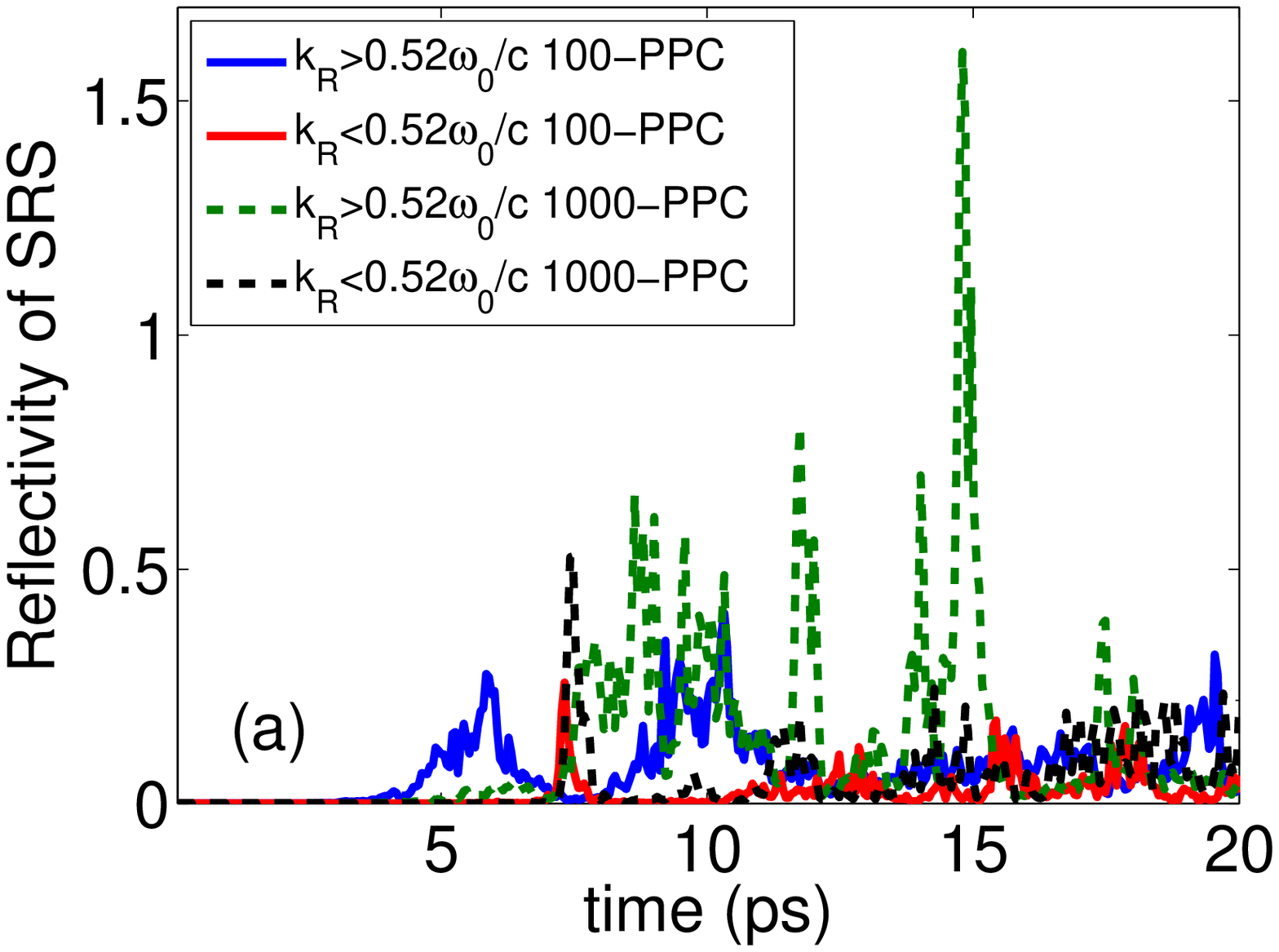}
\\\
\\\
\includegraphics[height=0.25\textwidth,width=0.4\textwidth,angle=0]{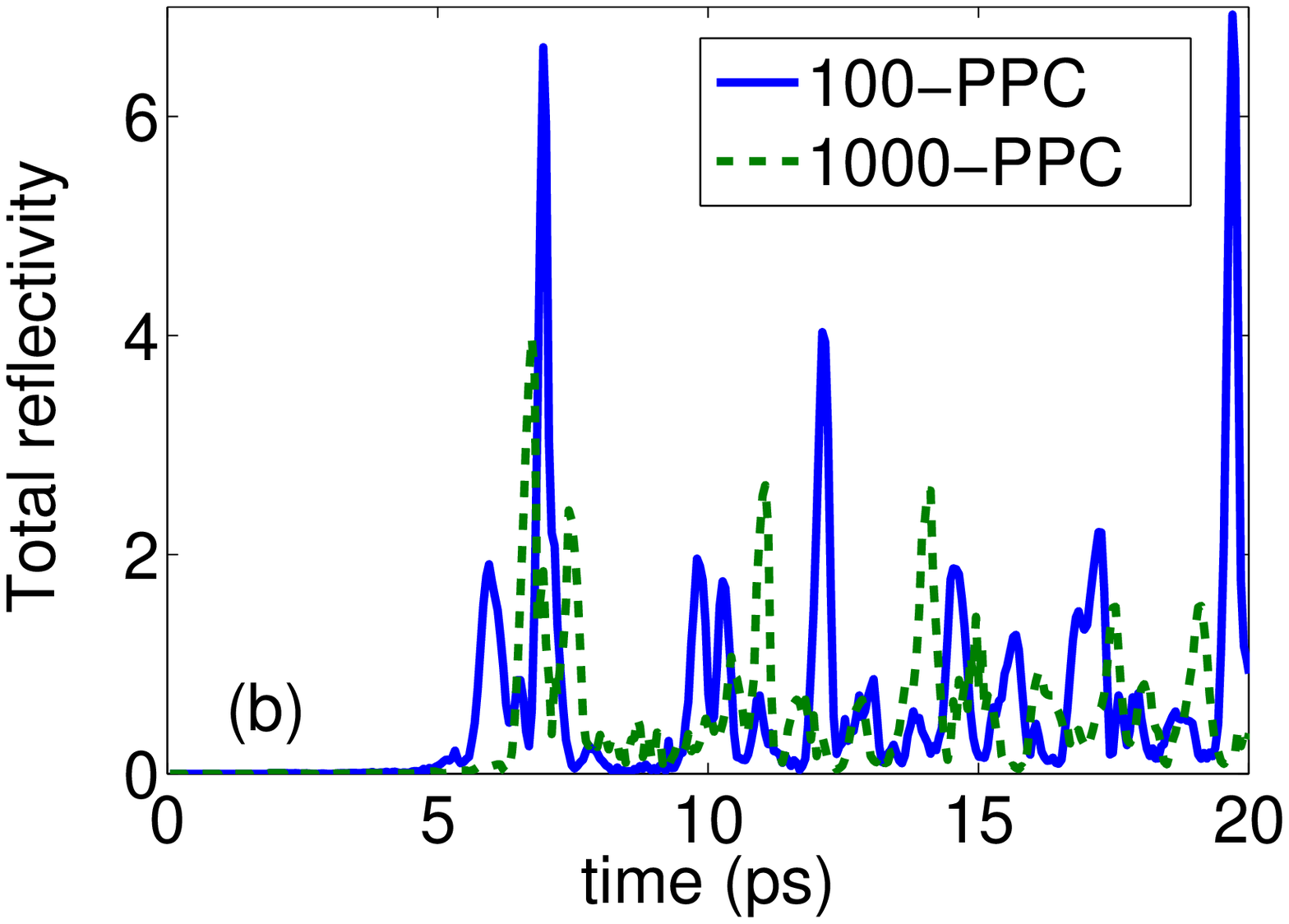}
\caption{(Color online) Evolution of (a) SRS reflectivity and (b)
total reflectiviy for the LT case. The incident laser intensity is
$5\times 10^{15}$W/cm$^2$ and the plasma flow is included.} \label{fig9}
\end{figure}

The larger reflectivities, especially the SRS reflectivities, in the
OSIRIS simulations compared to the HLIP simulations can be
attributed to many differences of the two codes. OSIRIS is fully
kinetic and nonlinear while HLIP lacks both. However, even in the
OSIRIS simulations, the dominant contribution to the SRS
reflectivity comes from the convective modes in the low density
region (see below). Therefore the seed levels for the convective SRS
and SBS can be important to their saturation and the resultant
reflectivities. To study the seed level effects, we repeat the LT
case with $I=5\times 10^{15}$W/cm$^2$ and plasma flow, by using
$1000$ particles per cell (PPC) for comparison. The results are
compared with the 100-PPC case as shown in Figs. \ref{fig9}(a) and
\ref{fig9}(b). For both cases, the SRS reflectivity is dominated by
modes of $k>0.52 \omega_0/c$, which are the convective modes in the
region of $n<0.2 n_c$ [Fig.\ref{fig9}(a)]. The time-averaged total
reflectivity drops from 64\% (100-PPC) to 50\% (1000-PPC)
[Fig.\ref{fig9}(b)]. This drop is mainly due to the drop of the SBS
reflectivity, which changes from 51\% (100-PPC) to 30\% (1000-PPC).
The SRS reflectivity increases from 13\%  (100-PPC) to 20\%
(1000-PPC) [Fig.\ref{fig9}(a)], due to competition between SRS and
SBS \cite{Berger98,Walsh84,Villeneuve87}. This also shows that in
the OSIRIS simulations, SRS and SBS are probably in the nonlinear
regime. In contrast, the level of SRS in the HLIP simulations is
much smaller.

\begin{figure}[htb!]
\includegraphics[height=0.55\textwidth,width=0.35\textwidth,angle=0]{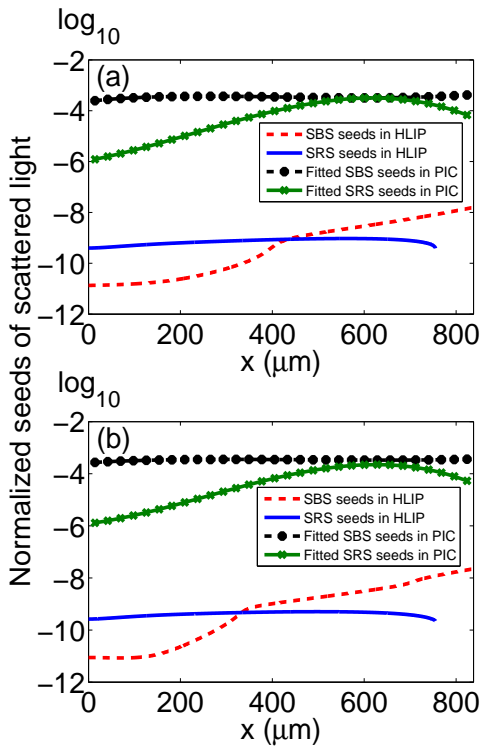}
\caption{(Color online) The normalized seed level profiles of all
SBS and SRS modes along the ray path for (a) the HT case and (b) the
LT case when $I=5\times 10^{15}$W/cm$^2$. All of the intensities are
normalized by $I$. The black dashed lines with sold circles show the
SBS seeds and the green sold lines with crosses are the SRS seeds
diagnosed in the OSIRIS simulations. The red dashed lines and blue
solid lines are the integrated total seeds for the SBS and SRS
backscattered light respectively given in HLIP based on the Thomson
scattering model \cite{Hao14}.} \label{fig10}
\end{figure}

To resolve the difference in reflectivity and pump depletion  between the fluid and
PIC simulations, we analyze the seed levels for the convective SBS
and SRS in the two kinds of simulations. In order to diagnose the
seed levels for the backscattered light in OSIRIS, the $B_z$ field
data is dumped every $20$ calculation steps and divided into
different windows in time and space. The time and spatial dimensions
of each such window are $4000 \Delta t$ and $5000 \Delta x$. Through
2D Fourier transform of the data in the time-space window, signals
of the backscattered light and incident light can be separated in
the $\omega$-$k$ phase space. Summing up the signals of the backward
light for SRS and SBS at different locations in the presence of the
pump, we can obtain the seed levels defined in the same way as those
used in HLIP.

\begin{figure}[htb!]
\includegraphics[height=0.23\textwidth,width=0.45\textwidth,angle=0]{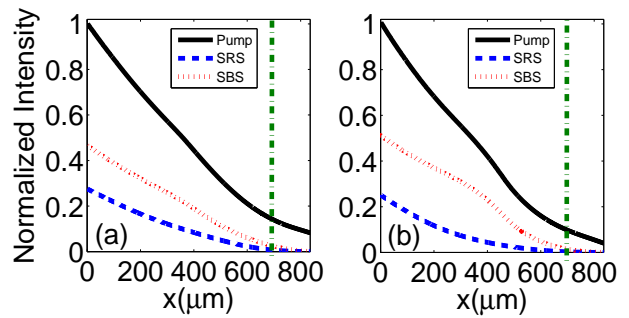}
\caption{(Color online) Spatial profiles of normalized intensities
of the pump and backscattered light for (a) the HT case and (b) the
LT case when $I=5\times 10^{15}$W/cm$^2$, by using the seeds fitted
from PIC simulations in HLIP code.} \label{fig11}
\end{figure}

Figure \ref{fig10} shows the seed levels in the two kinds of the
simulations for the HT and LT cases when $I=5\times
10^{15}$W/cm$^2$. The seeds in the OSIRIS simulations are many
orders of magnitudes larger than those calculated in HLIP. When HLIP
uses the fitted seed profiles from OSIRIS instead of its own seed
model, a significant growth of SRS can be obtained as shown in Fig.
\ref{fig11}. The effect of a higher seed level on SBS is less
significant, moderated by the pump depletions due to the inflated
SRS. Overall, the new HLIP simulations show a higher pump depletion
closer to the OSIRIS results. Other possible causes for difference
exist between the two codes, such as re-scattering of SRS and
generation of cavity \cite{Klimo10} (which is possible only at the
inflated SRS level and not possible at the level shown in the fluid
simulation with normal seed levels), kinetic and non-steady-state
physics all in OSIRIS but not HLIP. Another possibility is the
density-modulation-induced absolute SRS modes \cite{Nicholson76},
However,
the different seed levels should be an important 
cause for the difference in pump depletion between the OSIRIS and
HLIP simulation results.

\section{Discussion and Summary}
A recent PIC study of the temperature effects on LPI in SI also
found that the largest energy losses are due to the backscattering
from SBS rather than SRS when $T_e\leq5$keV and reflectivity of SBS
decreases as $T_e$ increases \cite{Weber15}, even though those
simulations were without flow and with a small scale length of 60
$\mu m$. In previous small-scale-length PIC simulations
\cite{Riconda11,Weber15,Weber12}, pump depletion through convective
SRS and SBS in the low density region was not significant, due to
the small lengths. The long-scale-length simulations here show that
the convective modes can lead to significant pump depletion before
the quarter-critical surface and an accurate assessment of this
requires control of the seed levels in simulations.

Neither of the two kinds of the simulations in the current work can
fully match for the backscattering measured in the
experiment\cite{Theobald12}. The overall pump reflectivities in the
experiment were much lower than the OSIRIS results, likely due to
the inflated seed levels. For the low intensities, the experiment
showed a stronger SBS reflectivity than SRS, agreeing with the HLIP
results. However, the experiments showed a rapidly increasing SRS
reflectivity as the spike beam intensity increased, eventually
exceeding the SBS reflectivity at $I=8\times 10^{15}$ W/cm$^2$. This
is very different from the HLIP results. One possible source of this
discrepancy is the hydro profile used in the simulations, which was
taken from the LILAC simulation that had only the drive beams, not
the spike beams. The profile may be significantly different from the
actual one when the spike beams were on. Furthermore, the SRS gain
model in HLIP does not include the possibility of absolute SRS and
high-frequency hybrid modes \cite{afeyan97} in the quarter-critical
region, which have been seen in previous 2D OSIRIS simulation in
$n=0.17-0.33 n_c$ with $I=2\times 10^{15}$ W/cm$^2$\cite{Yan14}. The
PIC simulations did show significant SRS reflectivity as the pump
intensity increased. However, the inflated seed levels for the
convective modes may have exaggerated the absolute levels.

In summary, SBS and SRS for typical laser and plasma conditions in
shock ignition have been studied using the fluid and PIC
simulations. Results show that SBS is the main cause for strong pump
depletion, which limits the intensity of laser light arrived at the
quarter critical density region. The plasma flow velocity gradient
is shown to affect the SBS reflectivity. The seed level analysis
also finds that the seed levels for both SRS and SBS in the PIC
simulations are much higher than those in an actual plasma, which
causes the stronger pump depletion in the PIC simulations.
Comparison with the experiment shows the need of new simulation
tools with both realistic seed levels and more comprehensive
physics.

\section{acknowledgments}
The authors would like to acknowledge the OSIRIS Consortium for the
use of OSIRIS. This work was supported by DOE under Grant No.
DE-FC02-04ER54789 and DE-SC0012316; by NSF under Grant No.
PHY-1314734; and by  National Natural Science Foundation of China
(NSFC) under Grant No. 11129503. The research used resources of the
National Energy Research Scientific Computing Center. The support of
DOE does not constitute an endorsement by DOE of the views expressed
in this paper.


\end{document}